\documentclass[preprint,showpacs,preprintnumbers,amsmath,amssymb]{revtex4}

\usepackage{graphicx}% Include figure files
\usepackage{dcolumn}% Align table columns on decimal point
\usepackage{bm}% bold math

\begin{document}

\preprint{}

\title{Interplay between quantum criticality and geometrical frustration in  Fe$_3$Mo$_3$N with stella quadrangula lattice}
%\title{The stella quadrangula lattice as a new geometrically frustrated system:
%Field- and impurity-induced ferromagnetism in Fe$_3$Mo$_3$N near the quantum critical point}

\author{Takeshi Waki}
\affiliation{Department of Materials Science and Engineering, Kyoto University, Kyoto 606-8501, Japan}
\author{Shinsuke Terazawa}
\affiliation{Department of Materials Science and Engineering, Kyoto University, Kyoto 606-8501, Japan}
\author{Teruo Yamazaki}
\affiliation{Department of Materials Science and Engineering, Kyoto University, Kyoto 606-8501, Japan}
\author{Yoshikazu Tabata}
\affiliation{Department of Materials Science and Engineering, Kyoto University, Kyoto 606-8501, Japan}
\author{Keisuke Sato}
\affiliation{Institute for Solid State Physics, The University of Tokyo, Kashiwa 277-8581, Japan}
%present address: Ibaraki National College of Technology, Department of Natural Science 866 Nakane, Hitachinaka 312-8508, Japan
\author{Akihiro Kondo}
\affiliation{Institute for Solid State Physics, The University of Tokyo, Kashiwa 277-8581, Japan}
\author{Koichi Kindo}
\affiliation{Institute for Solid State Physics, The University of Tokyo, Kashiwa 277-8581, Japan}
\author{Makoto Yokoyama}
\affiliation{Faculty of Science, Ibaraki University, Mito 310-8512, Japan}
\author{Yoshinori Takahashi}
\affiliation{Graduate School of Material Science, University of Hyogo, Koto Hyogo 678-1297, Japan}
\author{Hiroyuki Nakamura}
\affiliation{Department of Materials Science and Engineering, Kyoto University, Kyoto 606-8501, Japan}
\date{\today}

\begin{abstract}
In the $\eta$-carbide-type correlated-electron metal Fe$_{3}$Mo$_{3}$N, ferromagnetism is abruptly induced from a nonmagnetic non-Fermi-liquid ground state  either when a magnetic field ($\sim$14 T)
applied to it or when it is doped with a slight amount of impurity ($\sim$5\% Co).
We observed a peak in the paramagnetic neutron scattering intensity at finite wave vectors, revealing the presence of the antiferromagnetic (AF) correlation hidden in the magnetic measurements. 
It causes a new type of geometrical frustration in the \emph{stellla quadrangula} lattice of the Fe sublattice. 
We propose that the frustrated AF correlation suppresses the F correlation to its marginal point and is therfore
 responsible for the origin of the ferromagnetic (F) quantum critical behavior in pure Fe$_3$Mo$_3$N.

\end{abstract}

\maketitle
\begin{figure}[htbp]
\begin{center}
\includegraphics[width=0.7\linewidth]{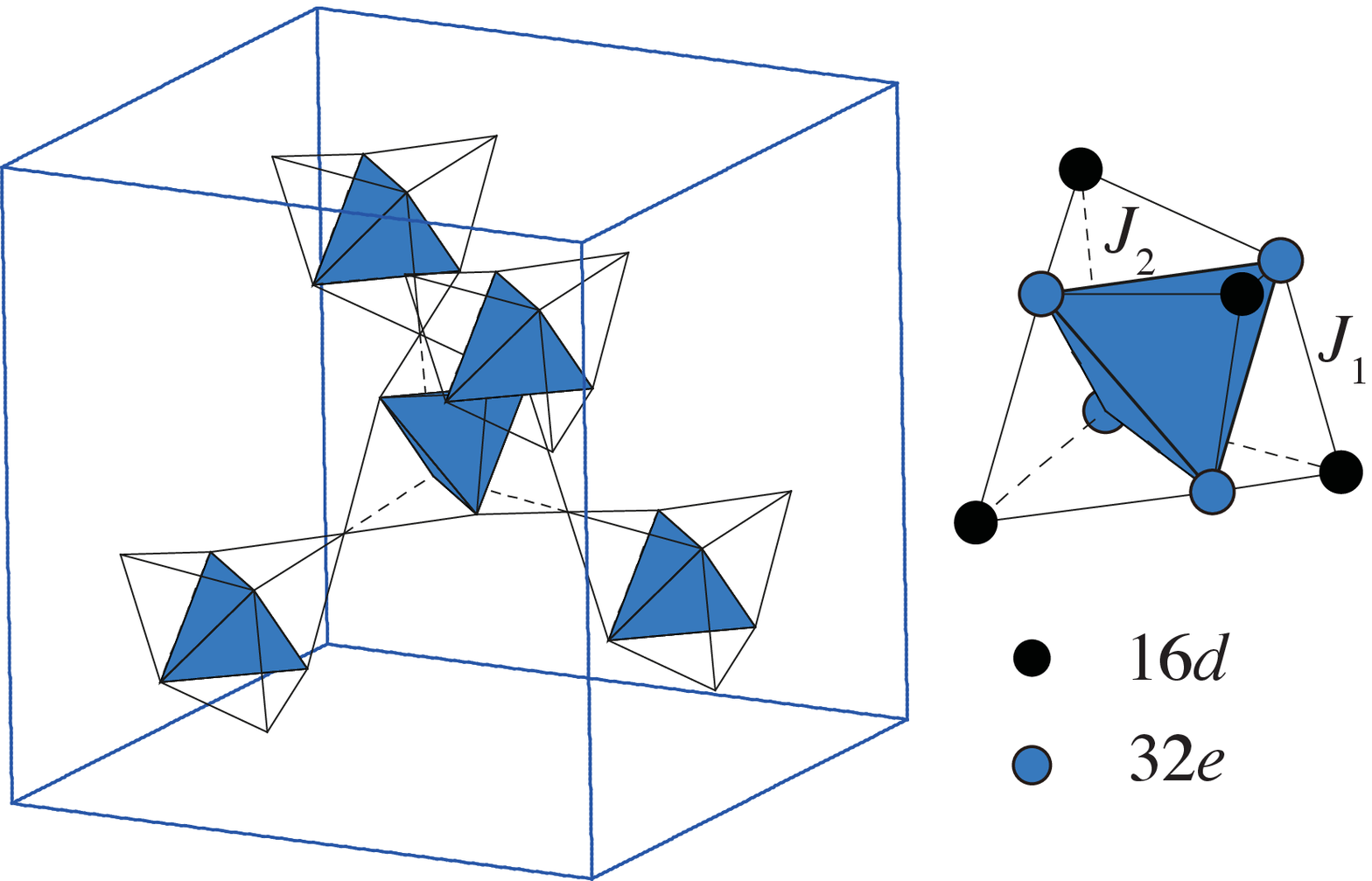}
\caption{\label{fig1}(color online) Schematic view of the \textit{stella quadrangla} lattice. The unit, a \textit{stella quadrangula}, consists of two nested regular tetrahedra with the same center of gravity.
}
\end{center}
\end{figure}

The geometric frustration in electronic degrees of freedom is a key factor responsible for the suppression of
long-range magnetic order that may lead to the realization of new and exotic quantum phenomena \cite{PhysicsToday}. 
Although the role of frustration was established in localized electron systems, no consensus has been achieved as far as their role in itinerant electron metals is concerned \cite{LacroixJPSJ2010}.
One of the reasons for this is the fact that a limited number of materials of this type are available.
Y(Sc)Mn$_{2}$, $\beta$-Mn, LiV$_{2}$O$_{4}$, etc.\ are some of the known examples  \cite{YScMn2}, whose electronic quasiparticle mass is commonly enhanced but
the manner in which it is enhanced is still under debate.
To resolve this issue, it is essential that we find other materials of a similar nature.
In this letter, we propose the $\eta$-carbide-type transition metal compound as a good candidate for testing geometric frustration in metallic magnets;
 the $\eta$-carbide-type structure (prototype: Fe$_3$W$_3$C, space group: $Fd\bar{3}m$)
  includes the {\it stella quadrangula} (SQ) lattice (Wyckoff positions $16d$ and $32e$) \cite{stellaquadrangula}, which is a corner-shared network of stellate (tetra-capped) tetrahedrons (see Fig.\ \ref{fig1}).
In an SQ as a unit, a small regular tetrahedron at $32e$ is nested in each $16d$ tetrahedron having the same center of gravity.
Furthermore, direct coupling is expected at the relatively short Fe-Fe bonds.
Even though $\eta$-carbide-type transition-metal compounds have been known for a long time as impurities in steel and refractory and hard materials, as well as promising candidates for catalysts, etc.\ \cite{carbides}, their quantum physical properties have been less studied thus far. 
(as an exception, see \cite{Sviridov}.)
% except a different approach to search strong ferromagnets \cite{Sviridov}.
We have recently reported that Fe$_3$Mo$_3$N with the $\eta$-carbide-type structure \cite{Bem} exhibits a non-Fermi liquid (NFL) behavior near the ferromagnetic (F) quantum critical point (QCP) under ambient pressure \cite{wakiFe3Mo3N}, which is different from the findings of literatures published previously \cite{Panda, Prior2}; $C/T$ ($C$:\ specific heat, $T$:\ temperature)
 shows a -$\log T$ divergence and reaches 128 mJ/(f.u.mol K$^2$) at 0.5 K,
and the resistivity shows a $T^{5/3}$ dependence.
Fe$_3$Mo$_3$N is a one of the ideal Fe-based materials exhibiting NFL behavior \cite{Brando}.
%We also found that another $\eta$-carbide-type compound Co$_3$Mo$_3$C shows an itinerant-electron metamagnetism (IEM) at $\sim$37 T \cite{wakiCo3Mo3C}.

The susceptibility $\chi$ of Fe$_3$Mo$_3$N obeys the Curie-Weiss (CW) law at high $T$ (effective moment $p_\mathrm{eff} = 2.14$ $\mu_\mathrm{B}$/Fe, Weiss temperature $\theta = 2$ K) and exhibits a broad maximum at $\sim$75 K \cite{wakiFe3Mo3N}.
Because the $\chi(T)$ maximum is commonly observed among exchange-enhanced Pauli paramagnetic metals undergoing itinerant-electron metamagnetism (IEM) \cite{IEMreview},
the application of a high magnetic field to Fe$_3$Mo$_3$N may yield interesting results.
IEM of Fe$_3$Mo$_3$N is also fascinating related with the recent interest in the phase diagram in the visinity of F-QCP and tuning of quantum criticality \cite{QCPmeta}.
While we have detected the IEM at $\sim$14 T, it is somewhat different from a typical IEM \cite{IEMreview}.
In order to determine the reason for this difference, impurity doping (Co for Fe) and inelastic neutron scattering measurements of Fe$_3$Mo$_3$N were performed.
In this letter, we discuss the characteristics of the SQ lattice as a new geometrically frustrated system,
 and we use this model to explain the absence of long-range order in pure Fe$_3$Mo$_3$N in spite of the strong electron correlation.

%\section{Experimental}
%synthesis
Polycrystalline samples of (Fe$_{1-x}$Co$_{x}$)$_{3}$Mo$_{3}$N, where $0 \le x \le 1$, were synthesized by a solid-state reaction of a mixture of transition metal oxides in a H$_{2}$-N$_{2}$ mixed gas stream \cite{Prior}.
Fe$_{2}$O$_{3}$, Co$_{3}$O$_{4}$, and MoO$_{3}$ were mixed in a molar ratio of $(1-x)/2:x/3:1$ and placed in a silica tube;
and the mixture was then fired in a gas stream of N$_{2}$ containing 10\% H$_{2}$ at 700$^{\circ}$C for 48 h followed by heat treatment at 1000$^{\circ}$C for 48 h.
In order to homogenize the samples, the heat treatment, interspersed with intermediate grinding, was repeated at least four times, resulting in a systematic variation in the 
magnetism against $x$, which is not in accordance with the literature data \cite{Prior2}.
%Prior and Battle have reported discontinuous appearance of $T_{C}$ against $x$ which may be due to the site preference of Co ion to $32e$ site \cite{Prior2}.
%To avoid site preference of Co ion, we performed heat treatment as high as possible and intermediate grindings at least 4 times.
%The obtained samples were checked by X-ray diffraction and revealed to be in single phases.
The magnetization $M$ at low fields was measured using a SQUID magnetometer, MPMS (Quantum Design) equipped in the LTM center, Kyoto University.
%Research Center for Low Temperature and Materials Sciences, Kyoto University.
%Magnetization
The high-field magnetization for field strengths up to 54 T was measured for pure Fe$_{3}$Mo$_{3}$N using a pulse magnet equipped in ISSP at 4.2--100 K.
Polycrystalline powder was filled into a cylindrical polyethylene tube that measured 6 mm in length and 2.5 mm in diameter.
For the neutron scattering experiments, approximately 20 g of the powder was packed into a vanadium tube with a diameter of 20 mm.
A triple-axis spectrometer ISSP-HER installed at the C1-1 cold neutron guide of the research reactor JRR-3M at Japan Atomic Energy Agency (JAEA), Tokai, was employed.
All measurements were performed at a fixed final wave vector $k_\textrm{f} = 1.45$ \AA$^{-1}$,
 with a collimation sequence of guide-$40^\prime$-open-open, and with a horizontal focusing analyzer.
The energy resolution was estimated to be 0.22 meV at zero energy transfer.

%\section{Results}

\begin{figure}[htbp]
\begin{center}
\includegraphics[width=0.85\linewidth]{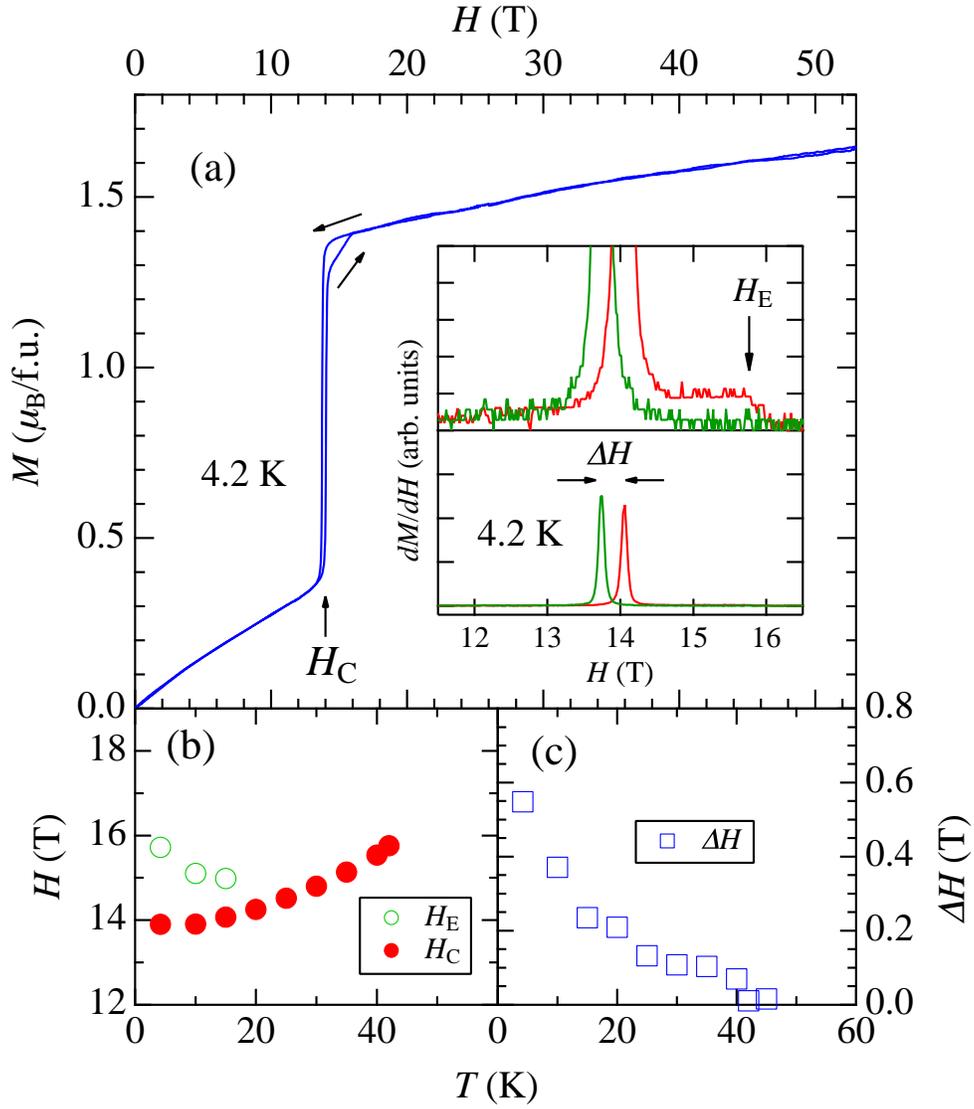}
\caption{\label{fig2} 
(color online) (a) High-field magnetization curve of Fe$_3$Mo$_3$N at 4.2 K.
The inset shows $dM/dH$ in both field-increasing and decreasing processes. 
The top half of the inset shows a magnified version of the image in the bottom half.
(b) $T$ dependences of the metamagnetic field $H_\mathrm{C}$ (average value of field-increasing and decreasing processes, closed circles) and
the field of the small anomaly in the field-increasing process, $H_\mathrm{E}$ (open circles).
(c) $T$ dependence of the hysteresis width $\Delta H$.
}
\end{center}
\end{figure}

Figure \ref{fig2} shows the results of the high-field magnetization measurements.
The signal from the pulse magnet corresponding to $dM/dH$ (at 4.2 K) is plotted against the external magnetic field $H$ in the inset of Fig.\ \ref{fig2}(a). 
$dM/dH$ shows diverging behavior at $\sim$14 T, corresponding to a sharp jump in the $M$-$H$ curve, which was obtained by integrating $dM/dH$. 
The metamagnetic fields $H_\mathrm{C}$ are different for the field-increasing and decreasing processes because of the first-order transition, 
although the difference (hysteresis width) $\Delta H$ is much smaller (0.56 T at 4.2 K) than that for a typical IEM \cite{IEMreview}.
$\Delta H$ decreases with $T$ and vanishes at $T_\mathrm{CM} \simeq 40$ K (Fig.\ \ref{fig2}(c)), which is the temperature at which
 the first-order IEM disappears, and corresponds to the substantial magnetic interaction in this compound.
%The critical field $H_\mathrm{C}$ at $T_\mathrm{CM}$ is estimated as 15.8 T.
The increase in $H_\mathrm{C}$ is roughly proportional to $T^2$ up to $T_\mathrm{CM}$ (Fig.\ \ref{fig2}(b) closed circles), as commonly observed in IEM \cite{IEMreview}.
%The magnetization jump at $H_\textrm{C}$ is reduced on approaching $T_\text{CM}$.
In the field-increasing process, a small step was observed at a field (defined as $H_\mathrm{E}$) slightly grater than $H_\mathrm{C}$.
$H_\mathrm{E}$ decreases gradually with $T$ and disappears at $\sim$20 K (Fig.\ \ref{fig2}(b) open circles).
The magnetization jump is sharp unlike that observed in other IEMs \cite{IEMreview} and rather similar to that of a spin flip in a localized spin system.
%This is in contrast to the fact that another $\eta$-carbide-type compound Co$_3$Mo$_3$C shows rather gradual IEM \cite{wakiCo3Mo3C}.
Successive magnetic transitions in the magnetic field are frequently observed in local spin systems;  
however, they are unlikely to be a part of a conventional IEM with uniform spin polarization, thus suggesting the presence of competing interactions or multiple order parameters.
%Observations of possible magnetovolume coupling and change in the lattice symmetry at $H_\mathrm{C}$ would be helpful to figure out the mechanism of the transition.

\begin{figure}[htbp]
\begin{center}
\includegraphics[width=0.80\linewidth]{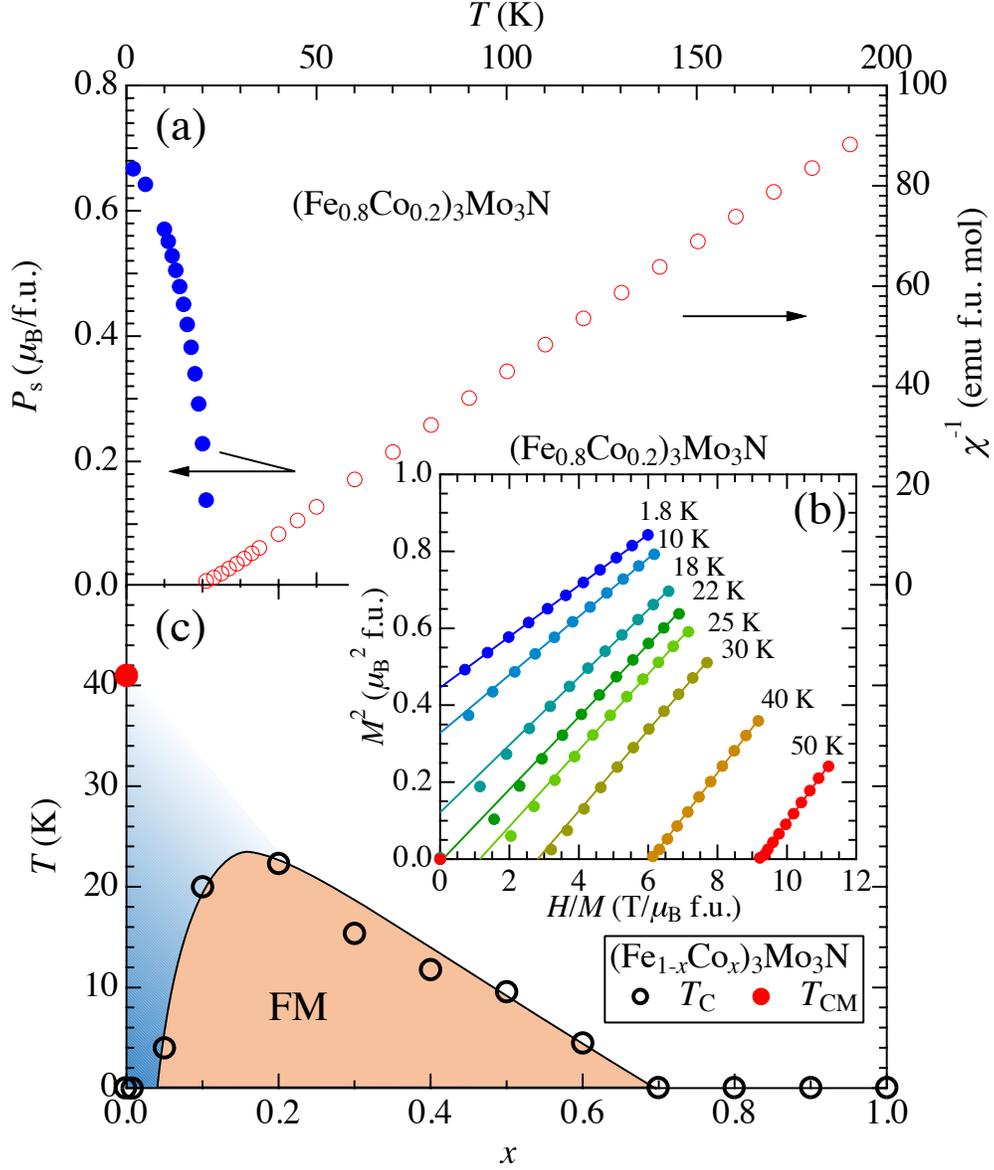}
\caption{\label{fig3}(color online) 
(a) $T$ dependences of inverse susceptibility and spontaneous magnetization (estimated from the Arrott plot) of (Fe$_{0.8}$Co$_{0.2}$)$_3$Mo$_3$N.
(b) Arrott plots of (Fe$_{0.8}$Co$_{0.2}$)$_3$Mo$_3$N.
(c) Magnetic phase diagram of the (Fe$_{1-x}$Co$_{x}$)$_3$Mo$_3$N system.
$T_\mathrm{C}$ (open circles) values were determined from  Arrott  plots. $T_\textrm{CM}$ (closed circle) is the hypothetical $T_\textrm{C}$ for pure Fe$_3$Mo$_3$N.
}
\end{center}
\end{figure}

The effect of impurity doping was investigated by replacing Co with Fe.
We have successfully synthesized a solid solution between Fe$_3$Mo$_3$N
 and the Pauli paramagnetic Co$_3$Mo$_3$N; the lattice parameter was found to vary linearly with the Co fraction $x$ (Vegard's law) in accordance with the literature data \cite{Prior2}.
As a typical example, the magnetic data of (Fe$_{0.8}$Co$_{0.2}$)$_3$Mo$_3$N are presented in Figs.\ \ref{fig3}(a) and (b).
At high $T$, $\chi$ follows the CW law well ($p_\mathrm{eff} = 2.55$ $\mu _\textrm{B}$/Fe, $\theta = 17.1$ K).
The Arrott plot ($M^2$ against $H/M$ plot) shows good linearity and intersects the vertical axis below 22 K, indicating the occurence of spontaneous magnetization $p_\mathrm{s}$ (=$0.28\:\mu _\textrm{B}$ at 5 K) below the Curie temperature $T_\mathrm{C} = 22$ K, despite the fact that both end compounds have nonmagnetic ground states.
The value of $p_\mathrm{eff}/p_\mathrm{s}$ is much larger than unity, as is characteristic of a typical weak  itinerant electron ferromagnet.
Similar analyses were carried out for other specimens to enable the construction of the magnetic phase diagram shown in Fig.\ \ref{fig3}(c) (detailed magnetic data  will be published elsewhere). 
The characteristics of this phase diagram are as follows.
A ferromagnetic phase appears by slight doping ($x < 0.05$).
With an increase in $x$, $T_\mathrm{C}$ increases rapidly up to a maximum at $x \sim 0.2$ and decreases almost linearly after the maximum up to $x \sim 0.7$.
It should be noted that the $T_\mathrm{CM}$ of pure Fe$_3$Mo$_3$N is much higher than the maximum value of $T_\mathrm{C}$ and is smoothly connected to the variation in $T_\mathrm{C}$ for $x \ge 0.2$.
In other words, long-range ferromagnetism appears to be suppressed particularly at $x \sim 0$, in spite of strong magnetic correlation.

\begin{figure}[htbp]
\begin{center}
\includegraphics[width=0.7\linewidth]{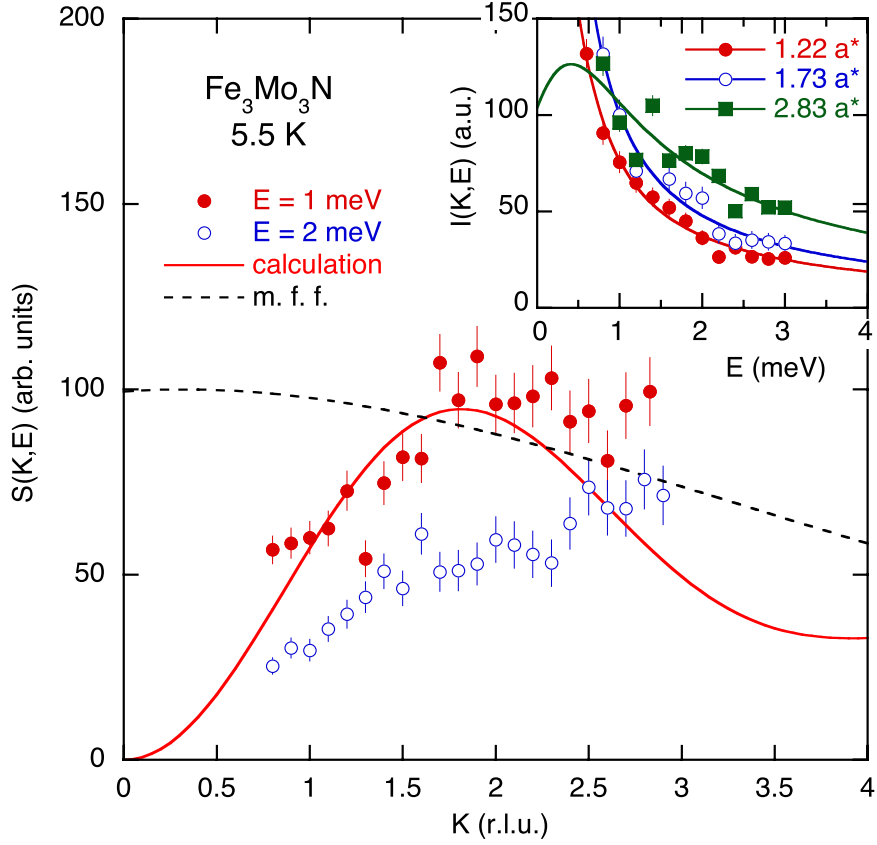}
\caption{\label{fig4}(color online) Wave number ($K$) dependences of quasi-elastic neutron scattering of Fe$_3$Mo$_3$N at 5.5 K for energy transfers of $E = 1$ and 2 meV.
$K$ is in the reciprocal lattice unit (r.l.u.).
The solid curve represents the scattering function $S(K, E)$ calculated for the SQ lattice (see text).
The broken curve represents the $K$ dependence of the squared magnetic form factor for Fe$^{3+}$.
The inset shows the $E$ scans measured at $K = 1.22$, 1.73, and 2.83 in r.l.u.
Solid curves represent the best fits to the Lorentzian quasi-elastic scattering function.
}
\end{center}
\end{figure}

To obtain information on the magnetic interactions in Fe$_3$Mo$_3$N, inelastic neutron scattering experiments were performed with the powder; unfortunately, single crystals have not been obtained.
Figure \ref{fig4} shows the wave number ($K$) dependences of quasi-elastic neutron scattering measured at 5.5 K.
Relatively strong scattering was observed, centered at finite but relatively small $K$, which roughly corresponds to the mean distances between Fe atoms, suggesting the presence of AF spin correlation together with the F correlation that was detected from macroscopic measurements.
The $E$-scan spectra, shown in the inset, were fitted with a Lorentzian quasi-elastic scattering function. 
The spectral widths, i.e., characteristic energies of spin fluctuations, are obtained as 0.4, 0.3 and 0.9 meV for $K = 1.22$, 1.78, and 2.83 in the reciprocal lattice unit, respectively.

\begin{figure}[tbh]  
\begin{center}  
\includegraphics[width=0.9\linewidth]{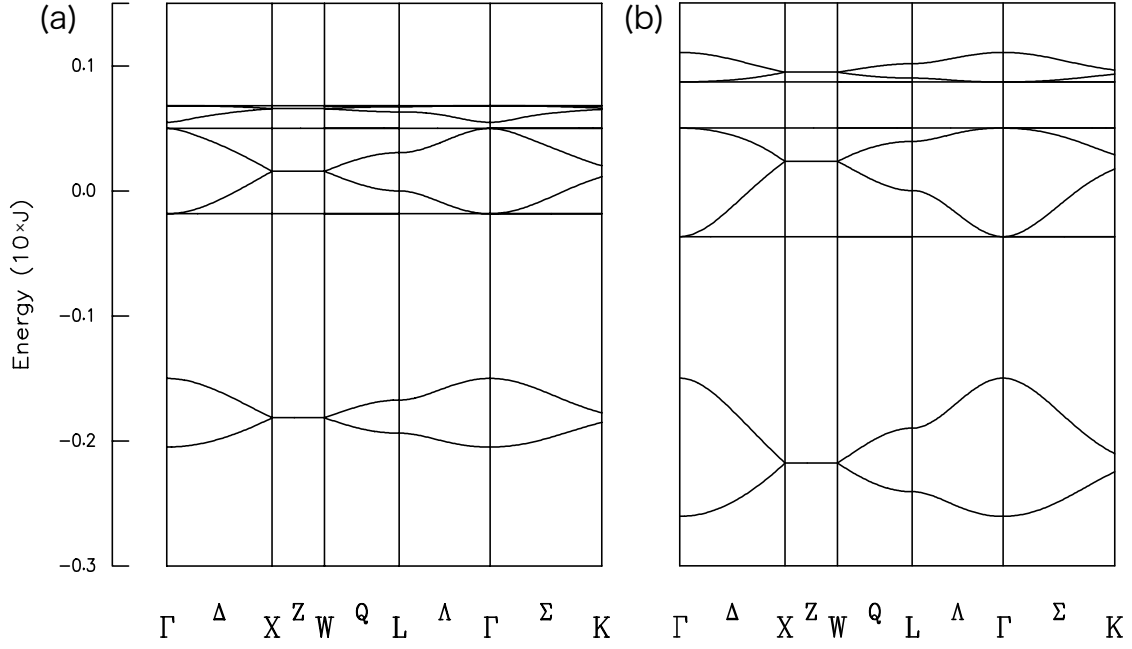}
\caption{\label{Jq}  Typical dispersion curves of $J(q)$ calculated for the SQ lattice with negative $J_2$.
(a) $J_2$-dominant ($|J_1|/J_2$ = $-0.5$) and (b) $J_1$-dominant ($-0.8$) cases. 
The results do not depend on the sign of $J_1$.
In these calculations, $z = 0.2937$ was used as the coordinate of $32e$, and the boundary of these states appears between $|J_1|/J_2$ = $-0.6$ and $-0.8$.}
\end{center} 
\end{figure}

Let us understand the geometrical characteristics of the SQ lattice on the basis of the Heisenberg model.
Although Fe$_3$Mo$_3$N appears to behave like an itinerant electron magnet, we believe that its spin density is relatively localized at the atomic site, as deduced from the strong electron correlation.
The SQ lattice consists of $16d$ and $32e$ sites arranged in the $Fd\bar{3}m$ space group. 
In actual $\eta$-carbide-type compounds, the nearest-neighbor (nn) and next-nearest-neighbor (nnn) distances are close and become exactly the same when $z = 0.3$, where $z$ is the coordinate of the $32e$ site $(z, z, z)$. 
Experimentally, $z = 0.2937$ \cite{Bem} and similar values \cite{Alconchel} have been reported for Fe$_3$Mo$_3$N, where the nn $16d$-$32e$ distance is 2.387 \AA\ and nnn $32e$-$32e$ distance is 2.549 \AA\ at room temperature.
Here, we consider only these nn and nnn interactions, namely, $J_1$ and $J_2$ interactions (see Fig.\ \ref{fig1}), respectively, and neglect the interactions of the neighbours located further along.
%Naively we expect $|J_1| \simeq |J_2|$ since nearly the same spin densities are expected at both $16d$ and $32e$ sites \cite{wakiFe3Mo3N}.
The signs of $J_1$ and $J_2$ are likely to vary because, empirically, both the F and the AF states appear in doped Fe-based $\eta$-carbide-type compounds \cite{Sviridov}. % , Fe6W6C}.
First, let us understand the nature of an isolated SQ.
In an extreme case, with $J_1 < 0$ and $J_2 = 0$, the SQ is not frustrated because the AF spins can be alternately assigned to the eight atoms.
Needless to say, the same is the case for $J_1 >0$ and $J_2 = 0$.
On the other hand, when $J_1 = 0$ and $J_2 < 0$, the isolated $32e$ tetrahedron is a typical frustrated unit.
These facts suggest that the SQ is frustrated only when $J_2$ is negative and dominant.
It should be noted that this is true not only when $J_1 < 0$ but also when $J_1 > 0$.
To verify whether this is true for an infinite SQ lattice, we calculated the dispersions of the Fourier transform of the exchange integral matrix $J(q)$  among 12 Fe atoms in a unit cell by assuming various $J_1/J_2$ ratios. 
Two typical results are shown in Fig.\ \ref{Jq}.
A flat dispersion is found along the highest branch when $J_1$ is not dominant (Fig.\ \ref{Jq}(a)), suggesting the degeneracies of the ground state, i.e., the presence of geometric frustration \cite{Harris}. 
On the other hand, $J(q)$ takes a maximum at the $\Gamma$ point, where the magnitude of $J_1$ is sufficiently larger than that of $J_2$ (Fig.\ \ref{Jq}(b)), indicating that $J_1$ tends to release the frustration. 
We confirmed that similar dispersions were obtained independent of the sign of $J_1$.
Thus, the SQ lattice is a unique frustrated system in which frustration is controlled by the $|J_1|/J_2$ ratio but not affected by the sign of $J_1$.

The anomalous magnetism of Fe$_3$Mo$_3$N can be easily understood on the basis of these characteristics.
The experimental results indicate the coexistence of the F and the AF interactions, which are formally denoted as  $J_1$ and $J_2$, respectively, as one of the simplest possibilities.
The suppression of long-range order in pure Fe$_3$Mo$_3$N is due to the frustration in the SQ lattice ($J_2$ dominant case).
In other words, the frustrated AF interaction suppresses the F correlation to its marginal point.
Metamagnetism can be interpreted as the transition to the $J_1$ dominant state, assisted by the external field.
The onset of static magnetism by slight doping is one of common features of a frustrated system.
It is not surprising that the F order is stabilized when the F interaction coexists with the AF interaction. 
We have tentatively calculated the spin correlation function $S(K, E)$ for the SQ lattice, assuming 
(1) $J_1 > 0$ and dominant $J_2 <0$,
(2) 2 up and 2 down spins in the $32e$ tetrahedron,
(3) an up spin at a $16d$ site when the neighboring $32e$ spins are 2 up and 1 down, and vice versa,
(4) no correlation between different SQs, and
(5) the magnetic form factor of Fe$^{3+}$.
Some of these assumptions may not be valid for the actual case, but have been made for the sake of simplicity.
$S(K, E)$ thus calculated and, furthermore, averaged over equal $K$ is illustrated
 by the solid curve in Fig.\ \ref{fig4}.
Although $S(K, E)$ should ideally be compared with the $E$ averaged values, there is reasonable agreement between the experimental results and the calculations carried out using our model, indicating that
 our model is consistent with the experimental results.
Although we have applied the localized moment scheme here, we have to reconcile the Heisenberg frustration and the electron itinerancy in further studies.

In summary, we found a steep metamagnetic transition in the $\eta$-carbide-type Fe-based compound Fe$_3$Mo$_3$N from the nonmagnetic NFL near F-QCP to a field-induced F state, and we established a magnetic phase diagram of the (Fe$_{1-x}$Co$_{x}$)$_{3}$Mo$_{3}$N system, where the F order is stabilized by a slight doping of the nonmagnetic element.
Neutron scattering measurements suggested that the AF and the F correlations coexist in Fe$_3$Mo$_3$N.
We proposed that the SQ lattice is a new geometrically frustrated system, in which the F and the AF interactions compete to select frustrated and non-frustrated states.
We applied this model to explain the suppression of long-range order in pure Fe$_3$Mo$_3$N.
In other words, the emergence of the F-QCP is interpreted as being the result of frustration.
Although a number of $\eta$-carbide-type transition-metal compounds are known to exist, their quantum physical properties have been less studied.
We believe that these compounds are good candidates for testing geometric frustration in metallic magnets and searching for new and exotic quantum phenomena.

%acknowledgment
\section*{Acknowledgement}
%We thank Prof. K. Yoshimura and Dr. H. Ohta for fruitful discussion and comments.
This study was supported by a Grant-in-Aid for Scientific Research on Priority Areas ``Novel States of Matter Induced by Frustration,"
a Grant-in-Aid for the Global COE Program, ``International Center for Integrated Research and Advanced Education in Materials Science,"
 and a Grant-in-Aid for Young Scientists (B) 21760531 from the Ministry of Education, Culture, Sports, Science and Technology of Japan.

\section*{Author Contributions}

T.W, S.T. and Y.Tabata conceived the experiments and analyzed the data. K.S, A.K. and K.K. provide valuable help on the pulse-high-field experiments and T.Y. and M.Y. on the neutron scattering experiment. Y.Takahashi made theoretical calculations and discussion. W.T. and H.N. co-wrote the paper. H.N. raid our and promoted all the works. All authors discussed the results and commented on the manuscript.

\section*{Competing financial interests}
The authors declare no competing financial interests.

\end{document}